# Multi-Scale Fiber Remodeling in HCM Using a Stress-Based Fiber Reorientation Law

**Mohammad Mehri[1], Hossein Sharifi[1], Kenneth S. Campbell [2], and Jonathan F. Wenk [1,3]**
[1] Department of Mechanical and Aerospace Engineering, University of Kentucky, Lexington, Kentucky, USA, {m.mehri@uky.edu, hossein.sharifi@uky.edu, jonathan.wenk@uky.edu}
[2] Division of Cardiovascular Medicine and Department of Physiology, University of Kentucky, Lexington, KY, USA, k.s.campbell@uky.edu
[3] Department of Surgery, University of Kentucky, Lexington, Kentucky, USA

**SUMMARY**

Quantifying fiber disarray, which is a prominent maladaptation associated with hypertrophic cardiomyopathy, remains critical to understanding the disease's complex pathophysiology. This study investigates the role of heterogeneous impairment of fiber contractility and fibrosis in the induction of disarray and their subsequent impact on cardiac pumping function. Fiber disarray is modeled via a stress-based fiber reorientation law within a multiscale finite element cardiac modeling framework called MyoFE. Using multiscale modeling capabilities, this study quantifies the distinct impacts of hypocontractility, hypercontractility and fibrosis on the development of fiber disarray and quantifies how their contributions affect the functional characteristics of the heart.

## 1 INTRODUCTION

Myocardial fiber disarray, a hallmark feature of hypertrophic cardiomyopathy (HCM), is defined as the irregular arrangement and disorganization of myocardial fibers. Heart muscle in HCM patients exhibits impaired contractility, fibrosis, and fiber disarray. The emergence of fiber disarray is commonly associated with contractile imbalance and fibrosis, yet the precise mechanisms are not fully understood. It has been reported that some HCM mutations result in hypercontractility, a condition previously observed to be caused by the mutation-induced unlocking of myosin heads from a state known as the super-relaxed state (SRX) [1]. HCM mutations do not all result in hypercontractility; different mutations alter the force generation mechanism in different ways, possibly including the inhibition of SRX [2]. The imbalance of contractile function (as a result of unequal fractions of hyper-hypo-contractile mutant proteins and wildtype proteins) will lead to cardiomyocyte disarray over time [3, 4]. In addition, replacement fibrosis, the most prevalent form of myocardial fibrosis in HCM [5], can induce heterogeneous changes in the material properties of muscle fibers [6] and is hypothesized to contribute to fiber disarray as well.

While previous modeling studies have incorporated the effects of fiber disarray on cardiac performance [7, 8], this study uniquely investigates the underlying mechanisms that trigger fiber reorientation and lead to myocardial fiber disarray. This study uses the MyoFE framework to identify the distinct contributions of contractile imbalance and fibrosis to fiber disarray. Moreover, this study aims to quantify the specific impact of microscopic abnormalities on global pump function. By elucidating how cell level alternation impacts organ level performance, this study aims to provide a deeper understanding of HCM disease progression.

## 2 METHODOLOGY

The current study presents a comprehensive modeling approach to investigate cardiac remodeling using a multiscale finite element (FE) framework, called MyoFE. This framework integrates various characteristics of cardiovascular function, including calcium handling, myofiber active contraction, and circulatory hemodynamics, in a closed loop model of the left ventricle (LV) and systemic circulation. In the present study, a fiber reorientation module is implemented within the MyoFE framework. At each time step, a novel stress-based reorientation law is used to compute fiber reorientation at each integration point in the mesh.

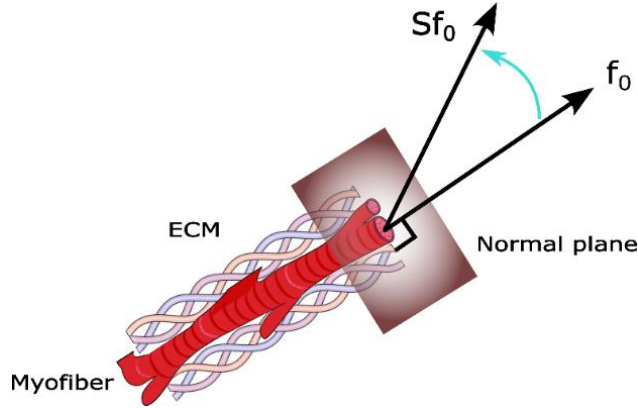

Figure 1: Stress based fiber reorientation. The unit vector $\boldsymbol{f_0}$ represents the initial direction of myofiber and collagen. $\boldsymbol{f_0}$ reorients toward the local traction vector $\boldsymbol{Sf_0}$, which is associated with the cross-sectional face of the fiber. $\boldsymbol{S}$ is the total stress tensor, encompassing both passive and active stresses.

As shown in Figure 1, $\boldsymbol{f_0}$ represents the unit vector along the fiber direction in the reference configuration, which is shared by the myofibers and collagen fibers. The traction vector ($\boldsymbol{Sf_0}$) is associated with the cross-sectional face of the fibers at a point in the LV. Our stress-based reorientation law incrementally reorients the fibers at each integration point in the LV mesh to be aligned with the traction vector as below:

$$\frac{d\boldsymbol{f_0}}{dt} = \frac{1}{\kappa}\left(\frac{\boldsymbol{Sf_0}}{\|\boldsymbol{Sf_0}\|} - \boldsymbol{f_0}\right) \quad (1)$$

where $\boldsymbol{S}$ is the 2$^{nd}$ Piola-Kirchhoff stress and $\kappa$ is a time constant that is employed to enforce the separation of time scales. More specifically, changes in fiber orientation occur over an extended period of time (e.g., days to weeks), whereas the current study achieves comparable fiber reorientation within several heartbeats. Through a gradual reorientation process, the difference between the fiber direction and the direction of traction experienced by the fibers is minimized. To account for the reorientation that occurs in both the myofibers and collagen, it is necessary to employ the total stress tensor, which encompasses both active and passive stress components.

Based on previous observations in myocardium with HCM, we implemented the following perturbations to our LV models. In some HCM mutations, a substantial variance in force generation is observed within cardiomyocytes at physiological calcium concentrations [3, 4]. In addition, disruption of the SRX state, which modulates the contraction of the heart, is also reported [3, 9]. Accordingly, for hypercontractile models we incorporated an increased (up to 100%) K$_{-SRX}$ value for 30% of cardiomyocytes distributed heterogeneously within the myocardium, where K$_{-SRX}$ is the transition rate of SRX in our contraction model called MyoSim [1, 9]. For the hypocontractile models, K$_{-SRX}$ is reduced with a similar percentage in cardiomyocytes with impaired contractility. In order to model the effects of heterogeneous replacement fibrosis in the myocardium, 30% of the cardiomyocytes are assigned with a 100% increase in stiffness, as well as being isotropic [5, 6, 8]. To facilitate head-to-head comparisons, a specific random distribution of abnormal cells (based on the number and location of cells) is assigned to all cases. For all models, the initial fiber configuration was assigned to the LV using a rule-based approach. Then, fiber reorientation was simulated until each case-specific fiber configuration reached steady state.

## 3 RESULTS AND CONCLUSIONS

As shown in Figure 2, all perturbed cases displayed considerable fiber disarray in contrast to the baseline model. As seen from the endocardial view, fiber disarray is more pronounced in hypocontractile and fibrous models whereas in the epicardial view, hypercontractility exhibits a greater degree of fiber disarray.

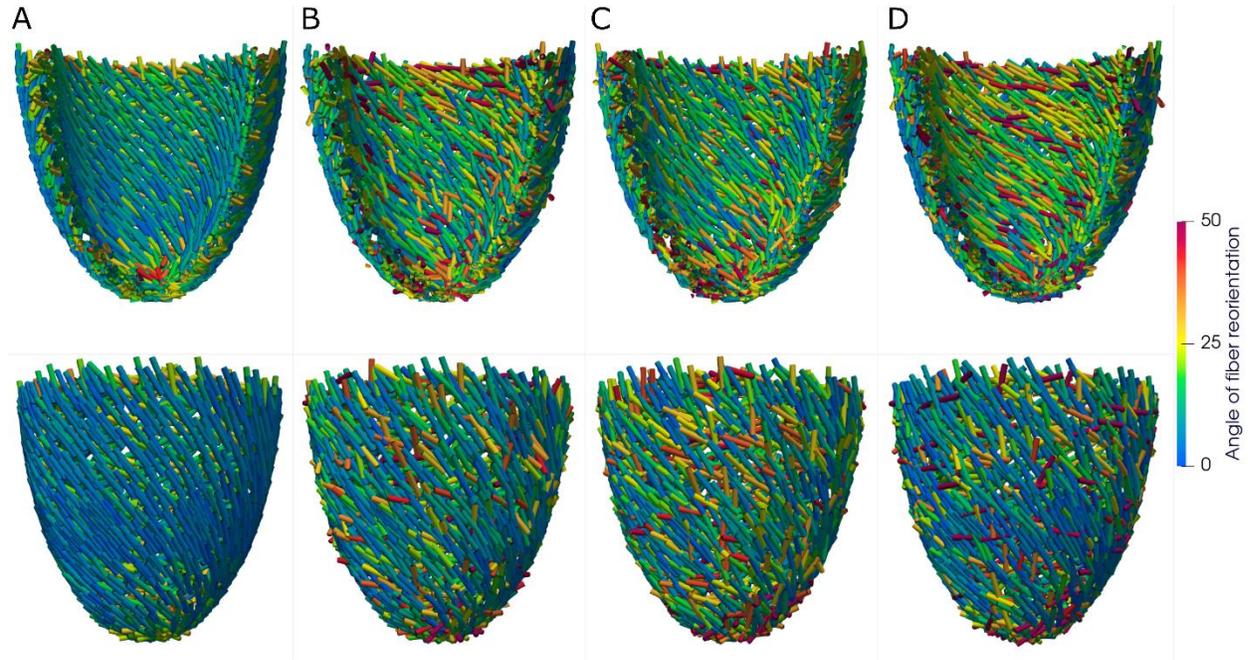

Figure 2: Final fiber configuration in LV models. A: Baseline LV model, B: LV model with heterogenous hypocontractility, C: LV model with heterogenous hypercontractility, D: LV model with heterogenous fibrosis. Top: Endocardial side of the myocardium. Bottom: Epicardial side of the myocardium.

In order to assess the sensitivity of fiber reorientation to the disruption of the SRX state, we implemented different levels of hypercontractility in order to evaluate the effect of $K_{-SRX}$ on the induction of fiber disarray. By increasing $K_{-SRX}$, the average value of fiber orientation within the LV was observed to increase proportionally as shown in Figure 3.

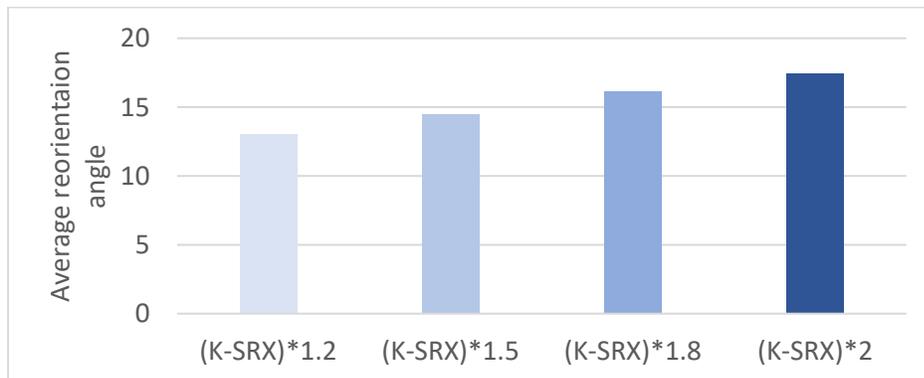

Figure 3: The average value of fiber reorientation in LV models with different increased values of $K_{-SRX}$ in hypercontractile cardiomyocytes.

As shown in Figure 4, pressure volume (PV) loops are compared to evaluate the effect of cell level perturbation on LV pumping performance. In the hypercontractile case, the end systolic volume (ESV) is reduced, and the PV loop is shifted to the left. Both hypocontractile and fibrous models result in a shrinkage of the PV loop and a significant increase in ESV due to reduced contractile abilities.

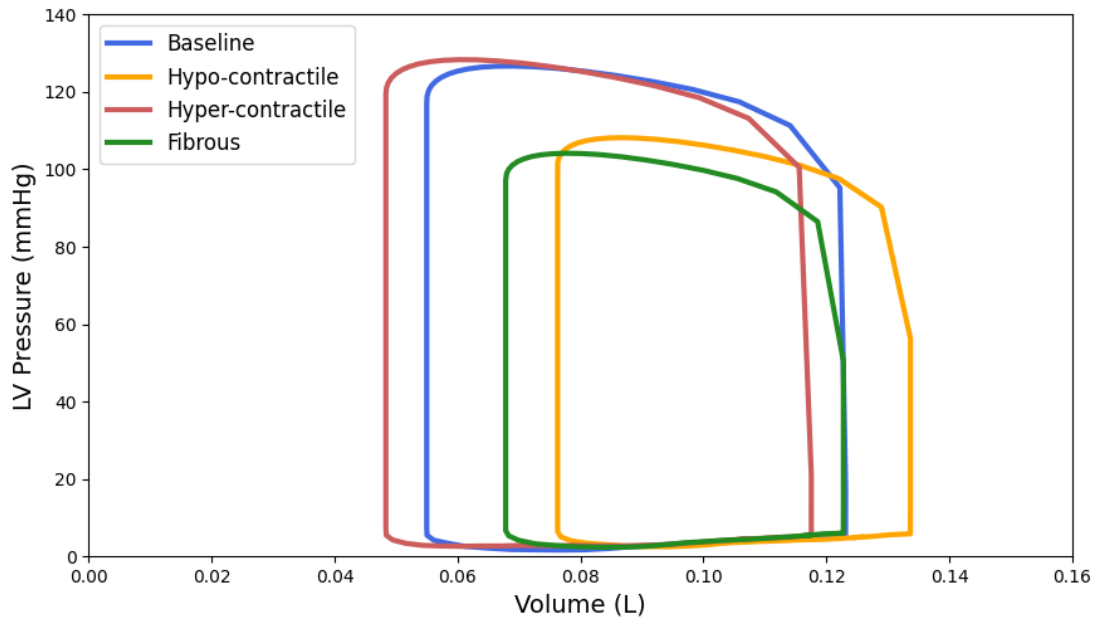

Figure 4: Pressure-volume (PV) loops of baseline and perturbed LV models.

In conclusion, the current study demonstrates the effectiveness of a stress-based law, within the multiscale cardiac modeling framework MyoFE, to predict fiber reorientation in the LV. These findings hold potential to enhance our understanding of pathological conditions and potentially aid in the development of therapeutic strategies to mitigate adverse remodeling effects due to HCM.